Numerical evaluation of deliberative discussions of the UK food system: stimuli, demographics, and opinion reversion

John Buckell, Nuffield Department for Population Health, Nuffield Department for Primary Health Care Sciences, University of Oxford.
Thomas Hancock, Institute for Transport Studies and Choice Modelling Centre, University of Leeds and Nuffield Department for Primary Health Care Sciences, University of Oxford.

**Abstract**

There is increasing acknowledgement – including from the UK government - of the benefit of employing deliberative processes (deliberative fora, citizens' juries, etc.). Evidence suggests that the qualitative reporting of deliberative fora are often unclear or imprecise. If this is the case, their value to policymakers could be diminished. In this study we develop numerical methods of deliberative processes to document people's preferences, as a complement to qualitative analysis. Data are taken from the Food Conversation, a nationwide public consultation on reformations of the food system comprising 345 members of the general public. Each participant attended 5 workshops, each with differing stimuli covering subtopics of the food system. In each workshop, individuals twice reported responsibility, from 0-10, for changing the food system for 5 stakeholders (governments, the food industry, supermarkets, farmers, individuals). Analyses examined individuals' perceptions of food system change responsibility. Governments were most responsible and farmers least so. We assessed variation by workshop content, and by demographics. Reported responsibility changed most for individuals, and changed least for the food industry. We devise a model to document a reversion effect, where shifts in perceptions on responsibility that occurred during workshops waned over time; this was strongest among those who intended to vote (rather than not to). These results can support qualitative analyses and inform food system policy development. These methods are readily adopted for any such deliberative process, allowing for statistical evaluation of whether they can induce opinion change.

**Introduction**

Improving methods for public involvement in policymaking, particularly from diverse backgrounds, is imperative for fair, democratic governance. There are fields of governance (e.g. climate change, food), where debate is so polarised that policymaking progress is intractable. Solutions backed by the public are more likely to be implemented and enforced, so it is important to engage the public in the policy-decision making process using methods that promote such engagement. As such, there is increasing acknowledgement – including from the UK government (CCC, 2022) - of the benefit of employing methods that allow the public to debate policy solutions and come to consensus, underpinned by robust evidence (Boulianne et al., 2018). Methods include deliberative fora, citizens' juries, and citizen assemblies. These approaches embody the idea of deliberation as a mechanism to resolve real-life problems, and emphasise convergence to solutions that are in the best interests of society rather than individuals. They allow participants time and resources to focus on one issue, consider it from multiple, often conflicting, positions, and recommend solutions. Through this, results of deliberative processes can be critical in policymaking and address "democratic deficits". Indeed, deliberative fora are increasingly being used in health and environmental policy research (Forde et al., 2024).

How deliberative methods are applied and integrated into policymaking decisions vary greatly. Though there is some generalised best practice guidance available (OECD, 2021), this currently falls short of providing specific insight into whether and how design features of a deliberative event influence participants' experiences and decision making. As deliberative methods are often more resource intensive than other participatory methods (e.g. surveys), improving our understanding about their design could lead to greater and more efficient use of these methods in policymaking.

Currently, analytical methods for public debates are almost all qualitative. Evidence suggests that the reporting of deliberative fora are often unclear or imprecise. If this is the case, their value to policymakers could be diminished. To ameliorate this issue, this research expands the set of quantitative methods for deliberative processes. Numerical information allows transparent evaluations of key elements of the deliberative process as a complement to qualitative analyses. This can provide new insights in the proceedings, serve as an aide in qualitative reporting of these events, promote greater replicability and research transparency, and may help to uncover researcher blind spots in findings.

Quantitative approaches to deliberative methods include deliberative polling (Fishkin, 2012), where participants' political opinions are recorded before and after debating a given issue, and any changes are measured. Relatedly, Reckers-Droog et al. (2020) use an extended Q-methodology approach where statements are ranked at two time points (rather than gave two opinion measures, as per deliberative polling). Further research has suggested the potential complementarity of public debates and choice experiments (Schoon and Chi, 2022). Indeed, the expansion of choice modelling methods can provide new ways to document public opinion (Mouter et al., 2021).

Here, we develop a new, quantitative complement to the qualitative approaches typical of deliberative events. We adapt the methods of stated choice experiments for preference elicitation (Hensher et al., 2015), using simple forms of data collection. We adapt choice modelling techniques (for an overview, see Hess and Daly, 2024), for an analytical framework to measure five aspects of opinion. First, to measure the response of opinions to various stimuli across a series of events that, in sum, comprise the deliberative event. Second, to measure variation across individuals' demographic characteristics. Since a range of information can be collected from respondents (e.g. demographics, behaviours, and attitudes), changes in opinion are linked to these. Moreover, attribution of changing opinions to either individual characteristics (and/or pre-existing beliefs/attitudes) or to deliberative fora events (and stimuli within those events) can be measured. Third, to measure opinion reversion, where within-event changes to people's views wane, to varying degrees, over time. Fourth, to measure multiple response outcomes and model responses simultaneously. The correlation between responses can, e.g. demonstrate who may be more susceptible to opinion change, or what views across different topics co-occur. Fifth, to capture variation over time, be that multiple observations per individual and/or calendar time across the deliberative event. This allows for testing the stability of opinions, and to map how different fora events or exogenous events that occur in between fora (e.g. the election of a new government) may influence opinions.

By design, our approach is a low-cost addition to deliberative methods (which are typically not low cost) and easy to apply. There is thus research productivity to be gained alongside scientific insight. We have provided all code in online repositories, and used open-source software to directly share our methods with other researchers, so that others can replicate in other settings; in turn generating further knowledge about the influence of design and context on deliberative events.

**Methods**

Setting and procedures

The Food Conversation was a nationwide public consultation on food system reform in the UK using deliberative fora (FFCC, 2024). 345 UK citizens in ten locations took part in a series of five workshops. Citizens were broadly representative of their location. They were invited to take part through a civic lottery by the Sortition Foundation (HVM, 2025). The events were carried out in four waves, each of 60-90 individuals, with the same procedure in each wave.

Within workshops, subject matter experts delivered a series of talks on four main topics (one per workshop; the final workshop was a synoptic and reflective session). Figure 1 shows the sequence and content of workshops. After the speakers, citizens had group-based discussions of challenges/solutions, mechanisms for change, and responsibility for change.

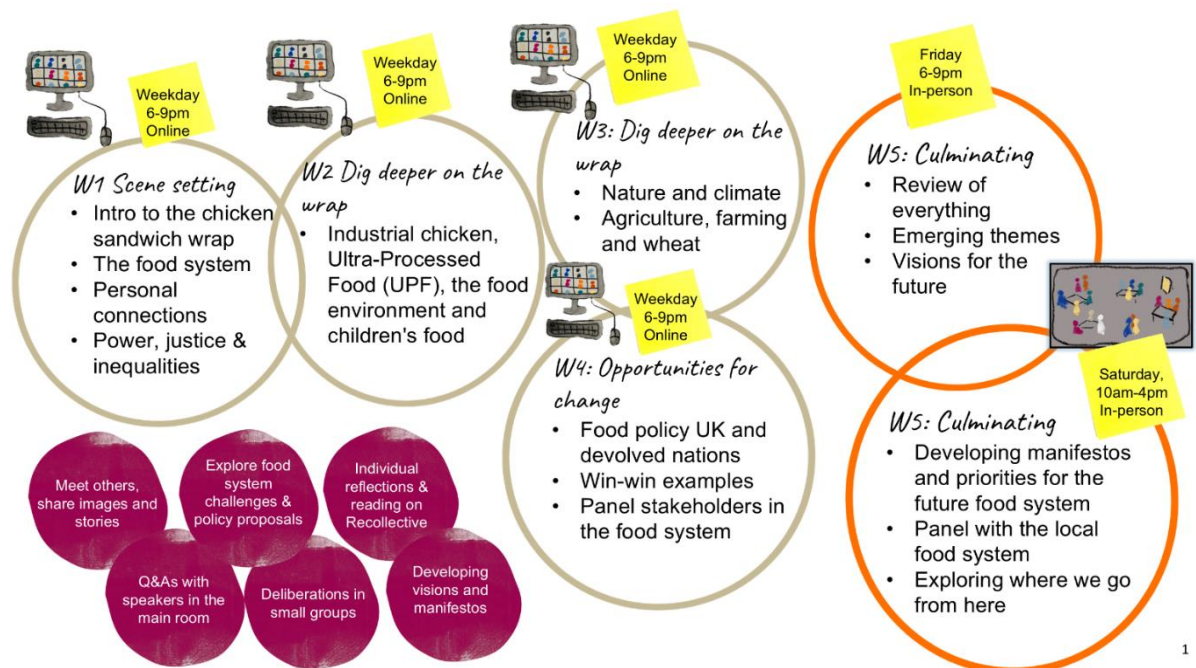

Figure 1: schematic of deliberative event order and content.

Data collection

Respondents were asked about responsibility for changing the food system. They submitted their perceived level of responsibility, on a scale from 0 to 10, for five stakeholders: governments, the food industry, supermarkets, farmers, individuals. These perceptions were recorded at the beginning and end of each workshop.

Other data collected were individuals' characteristics. This included age, gender, ethnicity, education, socioeconomic status, children in household, disability, rurality, and voting intention. Information on workshop content and order are naturally available, as are times and dates of workshops.

Choice modelling

The 5 separate ratings were entered by respondents on the 0-10 scale of food system change responsibility. A general model is of the form,

$$Rating = f(workshop) + f(wave) + f(demographics)$$

Which is operationalised as described below. (NB, rating can measured and modelled with any form of outcome: binary, ranking, multinomial, etc.)

Ratings are ordered outcomes, potentially correlated at the individual level, which are modelled using multivariate ordered logistic (logit) regressions estimate the probability of each response (i.e., on the 0-10 scale of food system change responsibility) for each observation in the data (i.e., individual at each time point). Five regressions are specified, one for each stakeholder. Thus we have:

$$P(Y_{i,s,t} = r) = \frac{e^{(\tau_{s,r} - V_{i,s,t})}}{1 + e^{(\tau_{s,r} - V_{i,s,t})}} - \frac{e^{(\tau_{s,r-1} - V_{i,s,t})}}{1 + e^{(\tau_{s,r-1} - V_{i,s,t})}} \quad (1)$$

$P(Y_{i,s,t_{i,t}} = r)$ is the probability that individual $i$ in time period $t$ reports that the level of responsibility for food system change for stakeholder $s$ is $r$ (which is one of 0-10; for example $r=8$ indicates the individual reporting the level of responsibility to be 8 on the scale of 0-10). Here, the index, $t$, denoting time period, refers to the sequence of the workshops and periods between them. There were five workshops, with data collected at the beginning and end of each workshop, hence the total periods, $T$, is 10. The waves occurred sequentially in calendar time; $t$ refers to the sequence of the deliberative events. (Calendar time is modelled separately – see below.) $\tau_{s,r}$ are stakeholder-specific threshold parameters estimated on the data, representing the threshold between each of the levels of responsibility on the scale. There are $R=11$ levels in the outcome, hence $R-1$ threshold parameters between them are estimated. $V_{i,s,t}$ is a stakeholder-specific linear predictor of the responsibility (see below). The functional form ensures that $\sum_R P(Y_{i,s,t} = r) = 1$.

The linear predictor comprises both stakeholder-specific and more generic components that enter all five linear predictors and are common across all five (i.e. do not vary across stakeholders).

$$V_{i,s,t} = U_{i,s,t} + \xi_{i,s} + \eta_{i,t} + \varepsilon_{i,s,t} \quad (2)$$

Where $U_{i,s,t}$ is the deterministic element of the linear predictor which varies at the level of each stakeholder equation,

$$U_{i,s,t} = \sum_{w=1}^{W} (\beta_{s,w} \cdot (I_w - \rho_i \cdot d_{i,w})) + \delta_{s,m} \cdot \chi_{i,m} + \Sigma_{z=1}^{Z} \gamma_{s,z} \cdot X_{i,z}. \quad (3)$$

Here, we have indicator variable $I_w$, which take a value of 1 if workshop $w$ has occurred (where in our case, these occur at $t \in \{2,4,6,8,10\}$, meaning e.g. for $t=5$, $I_1 = I_2 = 1$, $I_3 = I_4 = I_5 = 0$); $\beta_{s,w}$ measure the effect of the different workshops on the level of responsibility of the stakeholder. $\rho_i$ and $d_{i,w}$ together measure the reversion effect (see details below). $\delta_{s,m}$ are shifts to account for any differences across waves for the level of responsibility for a given stakeholder $s$, for wave $w$, that with $\chi_{i,m}$ an indicator variable taking a value of 1 if individual $i$ is in wave $m$ (in our case there are 2 wave parameters, with the third held constant for identification). Finally, $X_{i,z}$ is an individual characteristic ($Z=9$ characteristics in our data); $\gamma_{s,z}$ measure variation in the level of responsibility of the stakeholder across these characteristics.

During workshops, individuals' perceptions of food system change varied, and in differing magnitudes/directions across the five stakeholders. These effects were captured by the $\beta_{s,w}$ parameters in (3). Between workshops, these effects appeared to wane, and we call this a *reversion effect* where opinion change within workshops reverts towards its previous level. Since we measure both the levels of responsibility and the time between workshops, we can explicitly measure this reversion effect in our model. More specifically, we can measure two aspects of reversion. First, the overall level of reversion; that is, how much of the shift [in responsibility] from the workshop erodes over time. This is captured in $\rho_i$, which may vary across individuals. Second, we can measure the rate at which reversion occurs over the days following a workshop. This is captured in $d_{i,w}$. Further, we can measure deterministic variation in reversion (using individual characteristics) and random variation in reversion (with mixing distributions, which we approximate using draws).

$$\rho_i = \rho_{base} + \Sigma_{z=1}^{Z} \rho_z \cdot X_{i,z} + \sigma_\rho . draws_i^\rho \quad (4)$$

$\rho_{base}$ is a mean of the reversion effect to be estimated. When $\rho_{base} = 0$, there is no reversion and the level of responsibility is unchanged after the workshop; when $\rho_{base} = 1$, all the effect of the workshop has waned. Note that we do not impose $\rho_{base} > 0$ to allow for an opinion change during a workshop becoming stronger in between workshops (e.g. $\rho_{base} = -1$ implies that any change that occurs during a workshop is doubled in effect size). $\rho_z$ measures variation in the reversion effect across individual characteristics. $draws_i^\rho$ are draws from a standard normal distribution (at the level of the individual) with mean of zero and unit standard deviation, $draws_i^\rho \sim N(0,1)$; $\sigma_\rho$ is a standard deviation to be estimated, measuring unobserved, individual-specific variation in the reversion effect.

$$d_{i,w,m} = -1 + \exp\left(\frac{\alpha_i}{D}\right) . \exp\left(\frac{\alpha_i}{\Delta_{w,m} - D}\right) \quad (5)$$

Where $d_{i,w,m}$ is a decay function such that the overall reversion effect is in the range $[1-reversion_i, 1]$. $D$ is the range of days since the workshop (in our data, $D=17$) and $\Delta_{w,m}$ is the specific day after the workshop ($1 \leq \Delta_{w,m} \leq 17$), varying across waves but common across individuals within waves. $\alpha_i$ is an estimated set of parameters that dictate the rate at which the reversion sets in: higher values of $\alpha_i$ are quicker reversions.

$$\alpha_i = \alpha_{base} + \Sigma_{z=1}^{Z} \alpha_z \cdot X_{i,z} + \sigma_\alpha . draws_i^\alpha \quad (6)$$

$\alpha_{base}$ is a mean of alpha to be estimated. $z_i$ is a vector of individual characteristics; $\alpha_z$ measure how the rate of reversion varies across these characteristics. $draws_i^\alpha$ are draws from a standard normal distribution (at the level of the individual) with mean of zero and unit standard deviation, $draws_i^\alpha \sim N(0,1)$; $\sigma_\alpha$ is a standard deviation to be estimated, measuring unobserved, individual-specific variation in rate of reversion.

$\xi_{i,s}$ is a random effect capturing individual-specific, time-invariant unobserved variation, specifically:

$$\xi_{i,s} = \sigma_{\xi,s} . draws_{i,s}^\xi \quad (7)$$

$draws_{i,s}^\xi$ are draws from a standard normal distribution (at the level of the individual) with a mean of zero and unit standard deviation, $draws_{i,s}^\xi \sim N(0,1)$; $\sigma_s$ are stakeholder specific standard deviations to be estimated.

$\eta_{i,t,r}$ is a term entering all 5 equations, comprising an error component common across the stakeholders, capturing correlation across measures at the individual level (i.e., forming the multivariate model), and exogenous effects captured by calendar time.

$$\eta_{i,t} = \sigma_\eta \cdot draws_i^\eta + \delta_{calendar\ time} \cdot f(calendar\ time) \quad (8)$$

$draws_i$ are draws from a standard normal distribution (at the level of the individual) with mean of zero and unit standard deviation, $draws_i \sim N(0,1)$; $\rho$ is a standard deviation to be estimated. $\delta_{calendar\ time} \cdot f(calendar\ time)$ is general notation for temporal variation in responses (that is common across all five outcomes). This can be specified many ways (e.g. a linear trend, fixed effects, splines, etc.). In our model, we use calendar time fixed effects with *p-1* $\delta$ parameters, *p* being the number of time periods.

Models were built in a structured manner by testing components in a piecemeal way, refining the model at each step (by removing non-significant parameters, etc.) and then proceeding to the next component. For random model components, drawing from distributions was carried out using modified Latin hypercube sampling (Hess, 2006) with 5000 draws. Models were estimated using the Apollo package in R (Hess and Palma, 2019).

Robustness and sensitivity analysis

In preliminary testing, an ordered logit achieved a superior fit to a linear model and hence was adopted throughout.

A fixed effects specification of calendar time was preferred to a simple linear specification. A model in which calendar time effects were specified for each stakeholder model was discarded for two reasons. First, to reduce the number of estimated parameters. Second, due to multicollinearity given that waves are inherently related to calendar time.

**Results**

Sample

Data were available for 339 citizens of the 345 that participated in the events. For estimating models, we required individuals for whom data were recorded at the beginning and the end of workshops (to avoid confounding of between- and within-workshop events); 213 individuals met this standard. Table XXX presents the descriptive statistics for the sample. Of these, 63 were in wave 2, 65 in wave 3, and 85 in wave 4 of data collection. Waves 2 and 3 were in England and Wales, wave 4 was conducted in Scotland and Northern Ireland.

|  | Mean | N | % |
|---|---|---|---|
| Age | 50.15 |  |  |
| Age: missing |  | 53 | 24.9% |
|  |  |  |  |
| Gender: Female |  | 86 | 40.4% |
| Gender: Male |  | 71 | 33.3% |
| Gender: Non-binary |  | 2 | 0.9% |
| Gender: Missing |  | 54 | 25.4% |
| Ethnicity: Non-white | | 27 | 12.7% |
| Ethnicity: White | | 133 | 62.4% |
| Ethnicity: Missing | | 53 | 24.9% |
| | | | |
| Education: Higher | | 84 | 39.4% |
| Education: Not higher | | 76 | 35.7% |
| Education: Missing | | 53 | 24.9% |
| | | | |
| IMD category: 1 | | 15 | 7.0% |
| IMD category: 2 | | 40 | 18.8% |
| IMD category: 3 | | 26 | 12.2% |
| IMD category: 4 | | 22 | 10.3% |
| IMD category: 5 | | 15 | 7.0% |
| IMD category: Missing | | 95 | 44.6% |
| | | | |
| Children in household: Yes | | 43 | 20.2% |
| Children in household: No | | 117 | 54.9% |
| Children in household: Missing | | 53 | 24.9% |
| | | | |
| Disability: Yes | | 36 | 16.9% |
| Disability: No | | 124 | 58.2% |
| Disability: Missing | | 53 | 24.9% |
| | | | |
| Location: Urban | | 78 | 36.6% |
| Location: Rural | | 40 | 18.8% |
| Location: Missing | | 95 | 44.6% |
| | | | |
| Voting intention: Right | | 26 | 12.2% |
| Voting intention: Left | | 70 | 32.9% |
| Voting intention: Abstain | | 53 | 24.9% |
| Voting intention: Missing | | 64 | 30.0% |

Table 1: Descriptive statistics for analytical sample. IMD: Index of Multiple Deprivation in quintiles (1 most deprived; 5 least deprived).

Responsibility for Food System Change: Descriptive statistics

| | Responsibility for Food System Change | | | | |
|---|---|---|---|---|---|
| | Pooled Mean | Mean at t=1 | Mean at t=10 | Change | p-value |
| All | 7.47 | 7.28 | 7.78 | 0.50 | 0.002 |
| | | | | | |
| Government | 9.03 | 8.68 | 9.37 | 0.69 | 0.001 |
| Supermarkets | 8.14 | 7.50 | 8.21 | 0.71 | 0.017 |
| The Food Industry | 7.96 | 8.04 | 8.16 | 0.12 | 0.065 |
| Farmers | 6.08 | 6.81 | 6.07 | -0.74 | 0.047 |
| Individuals | 6.14 | 5.34 | 7.07 | 1.73 | 0.000 |

Table 2: Food system change responsibility. All - all five stakeholders; Mean – the mean of the response in the sample; Overall mean – mean of response pooled across all 10 time periods; Mean at t=1 – mean of response at the beginning of workshop 1; Mean at t=10 – mean of response at the end of workshop 5; Change – difference in response across the workshops; p-value – from paired t-test of the change.

Table 2 reports results for the measures of food system change responsibility in the sample. The pooled mean reported responsibility across both stakeholders and time periods was 7.47. Variation across stakeholders can be seen. Across all periods, governments were seen to be most responsible for food system change, with a pooled mean of 9.03, and farmers were seen to be least responsible, with a pooled mean of 6.08. Individuals' responsibility was very close to that of farmers, at 6.14. Supermarkets and the food industry were viewed to be more responsible than individuals and farmers, but less responsible than governments, with pooled means of 8.14 and 7.96, respectively.

The level of responsibility varied over the deliberative event in direction and magnitude across stakeholders. In three cases - governments, supermarkets, and individuals - responsibility increased over the deliberative event. For the food industry, no change in responsibility was observed. For farmers, responsibility declined over the deliberative event.

Figure 2 illustrates these findings and provides further information. The vertical alignment of the scores across each pane shows differences in the responsibilities of stakeholders. The gradient of each pane from t=1 to t=10 shows the overall change. The graph shows that the level of responsibility also varies considerably both within (blue shading) and between (pink shading) workshops. It also shows that responses to workshops vary across stakeholders: workshop 1 had positive and negative effects on responsibility, depending on the stakeholder. The amount of variation also changed: responsibility for farmers varies to a greater degree than responsibility of the food industry.

Responsibility for Food System Change: Choice modelling

Table 3 presents the abridged model estimates from the multivariate analysis of food system change responsibility (full results in Appendix XXX).

Workshop effects

The *Workshop effects* rows of Table 3 show associations of responsibility measures and each of the workshops. For governments, statistically significant positive associations were found for workshop 2 (industrial chicken, ultra-processed food, the food environment, and children's food; beta=0.94; robust t-ratio: 4.14) and for workshop 4 (UK food policy, win-win examples, panel of stakeholders; beta=0.39). For supermarkets, statistically significant positive associations were found for workshop 1 (the food system, personal connections, and power/justice/inequalities; beta=1.56; robust t-ratio: 5.79) and for workshop 4 (UK food policy, win-win examples, panel of stakeholders; beta=0.73; robust t-ratio: 3.67). Statistically significant negative associations were found for workshop 2 (industrial chicken, ultra-processed food, the food environment, and children's food; beta=-0.82; robust t-ratio: -4.27). For the Food Industry, no statistically significant associations were found. For farmers, a statistically significant positive association was found for workshop 3 (nature/climate, agriculture/farming/wheat; beta=0.89; robust t-ratio: 4.83), and a statistically significant negative association was found for workshop 1 (the food system, personal connections, and power/justice/inequalities; beta=-1.56; robust t-ratio: -5.98). For Individuals, statistically significant positive associations were found for workshops 3 (beta=0.37; robust t-ratio: 2.41), 4 (beta=0.67; robust t-ratio: 3.84), and 5 (beta=0.76; robust t-ratio: 3.14).

Demographic effects

The *Demographics* rows of Table 3 show associations of responsibility measures and individuals' characteristics. Female respondents believe that individuals were more responsible for food system change than male respondents (beta=0.96). Those who live in rural areas believe that farmers were less responsible for food system change than those who live in urban areas (beta=-1.79).

Reversion

Table 3's two right-hand columns, *Reversion* and *Alpha*, together with Figure 3, describe the reversion effects. The reversion parameters measure the extent to which opinion changes that occurred during a workshop reverts back to its starting point. When reversion=1, all of the opinion change in a workshop has waned. When reversion=0, all of the opinion change in a workshop has remained. When reversion>1, any opinion change during the workshop is enhanced. The parameters retained in the model are base, rural (versus urban) and did not vote (versus did vote). From these, four groups (with respective reversions) emerge as combinations of these parameters – rural non-voters (0.72), rural voters (0.33), urban non-voters (1.07), and urban voters (0.68). These are displayed in Figure 3. The Alpha parameters govern how quickly reversion sets in, and this varies across demographics. For higher values of alpha, reversion sets in more quickly, which can be seen in Figure 3. Rural non-voters and urban voters have approximately the same reversion (around 70% of the change in the workshop is retained). Urban voters' opinion reverts to its floor after 8 days; whereas rural non-voters' opinion reverts to its floor after 13 days.

Other model features

Some evidence was found in support of other model features, though since these are not a primary focus of our analysis, are only briefly discussed. Differences across waves were found in some, but not all, cases. Some evidence for random heterogeneity for stakeholders was found (i.e. sigma parameters in eqn. (7)). No evidence was found for random heterogeneity in either reversion nor alpha, that is, we did not find evidence to reject $\sigma_{rev} = 0$ in eqn. (4) nor $\sigma_{alpha} = 0$ in eqn. (5). We found evidence in support of the random effect in the error component, that is we rejected $\rho = 0$ in eqn. (8). This indicates that some individuals generally give higher scores than others.

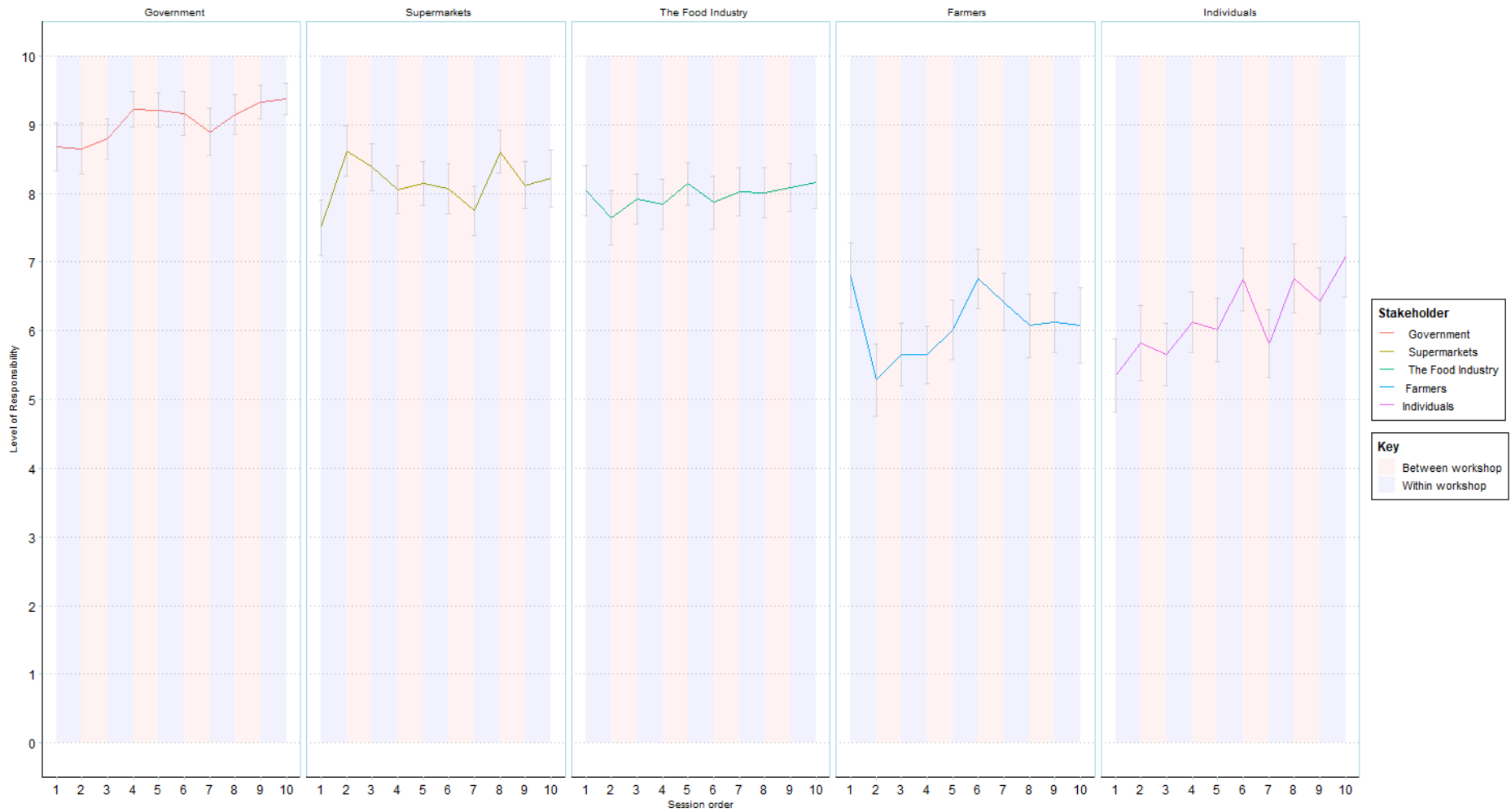

Figure 2: Food system change responsibility over the sequence of workshops of the deliberative event, by stakeholder. Time index, $t$, is on the x-axis. Mean (across respondents) levels of responsibility are on the y-axis. Data were recorded at the beginning and end of each workshop, hence blue shading is within workshops and pink shading is between workshops.

|  | Government | | Supermarkets | | The Food Industry | | Farmers | | Individuals | | Reversion | | Alpha | |
|---|---|---|---|---|---|---|---|---|---|---|---|---|---|---|
|  | Estimate | Rob.t-ratio(0) | Estimate | Rob.t-ratio(0) | Estimate | Rob.t-ratio(0) | Estimate | Rob.t-ratio(0) | Estimate | Rob.t-ratio(0) | Estimate | Rob.t-ratio(0) | Estimate | Rob.t-ratio(0) |
| *Workshop effects* | | | | | | | | | | | | | | |
| Workshop 1 | 0.13 | 0.53 | 1.56 | 5.79 | -0.45 | -1.56 | -1.56 | -5.98 | 0.46 | 1.79 | | | | |
| Workshop 2 | 0.94 | 4.14 | -0.82 | -4.27 | 0.16 | 0.92 | 0.33 | 1.93 | 0.26 | 1.59 | | | | |
| Workshop 3 | -0.54 | -0.29 | -0.22 | -1.28 | 0.68 | 0.42 | 0.89 | 4.83 | 0.37 | 2.41 | | | | |
| Workshop 4 | 0.39 | 1.97 | 0.73 | 3.67 | 0.25 | 1.59 | -0.35 | -1.79 | 0.67 | 3.84 | | | | |
| Workshop 5 | 0.20 | 0.76 | -0.15 | -0.43 | -0.26 | -0.12 | 0.14 | 0.59 | 0.76 | 3.14 | | | | |
| *Demographics* | | | | | | | | | | | | | | |
| Base | | | | | | | | | | | 0.33 | 3.56 | 119.17 | 1.78 |
| Female | | | | | | | | | 0.96 | 2.94 | | | | |
| Ethnicity: non white | | | | | | | | | | | | | -112.69 | -1.98 |
| Higher education | | | | | | | | | | | | | 83.67 | 1.42 |
| IMD over 7 | | | | | | | | | | | | | | |
| Rural | | | | | | | | -1.79 | -3.73 | | 0.35 | 1.18 | -9.85 | -1.57 |
| Voting: Right | | | | | | | | | | | | | | |
| Voting: did not vote | | | | | | | | | | | -0.39 | -2.98 | | |

Table 3: Model results from multivariate analysis of food system change responsibility (abridged; full results in Appendix XXX). Estimate – parameter estimate, Rob t-ratio(0) – robust t-ratio versus 0.

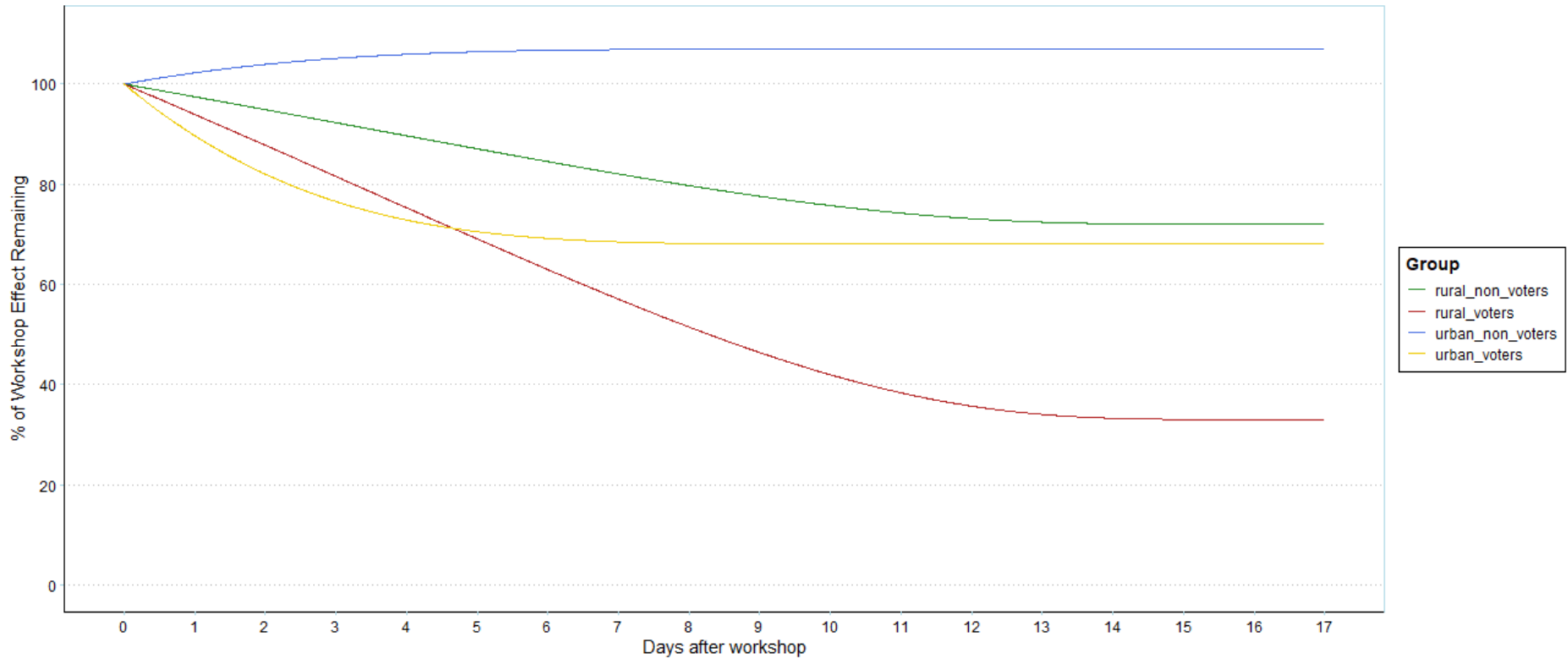

Figure 3: Reversion effects by group. Days following the workshop, $\Delta_{i,wm,}$, is on the x-axis. The % of workshop effect remaining is on the y-axis. Each plotted path takes the estimated reversion and alpha parameters from the model and applies them for the groups identified by those parameters. That is, the plotted paths show the reversion effect by group. When y=100, at the end of the workshop, no reversion has taken place and all of the workshop effect on perceived level of responsibility remains. When y=0, all of the workshop effect has waned and the perceived level of responsibility is as it was prior to the workshop. For example, "rural non-voters" (green line) are those people who reported living in a rural area and abstaining from voting; their opinion reverts over time, to 72% of its workshop shift. For y>100, as in urban non-voters, there has been a boost and the workshop effect is amplified over time.

**Discussion**

*Summary of findings*

We have set out a numerical complement to qualitative analysis of deliberative events. Data collection of food system change responsibility scales, and information on individuals, allowed analyses of these. We developed a multilevel, multivariate ordered logit analysis of five outcomes over the course of the deliberative event. Our main analyses focussed on measuring opinion change within the event's workshops, variation in opinion across sociodemographic characteristics, and a reversion effect where opinion change during the workshops eroded over time.

*Comparisons to qualitative findings in the Food Conversation*

Our quantitative results align, where comparable, to qualitative findings of the Food Conversation (cf. HVM, 2025; FFCC, 2024). The main finding on food system change responsibility was governments bore greatest responsibility. This accords with our finding that the highest recorded level of responsibility was for the government. There was a further sense of increasing onus on governments to make change. This was reflected in the increased score of government responsibility over the deliberative event. In the reports, there was also a sense that both supermarkets and the food industry could use their considerable power to effect change for good in the food system. Again, this accords with our finding of them scoring higher than farmers and individuals, but lower than governments. There was a finding that, as citizens learned more about the food system, there was a growing sense that farmers were treated unfairly by the food system. This is in keeping with the finding that, over the course of the deliberative event, the measured level of responsibility for farmers declined. Finally, there were an emergent theme of community power: that citizens could rally to foster positive change in the food system. This aligns with the increase in the measured level of responsibility for individuals.

*Comparison to prior literature*

Previous research uses choice modelling methods, termed participatory value evaluation (PVE) to document public opinion (Mouter et al., 2021). These are static assessments of opinion, and are not measured in response to either workshop content or to group discussion. In contrast to this study, however, it is easier to gather (a) large samples, and (b) complete sets of demographic data in surveys. Q-methodology is being used increasingly in health (Churruka et al., 2021). As with PVE, Q-methodology typically records responses at a single time point. Reckers-Droog et al. (2020) extend this, where various statements were ranked by 24 individuals at two time points. While similar to our approach, this method uses several outcomes and a factor analysis to find commonalities amongst them. A similar approach, using a latent variable, was used in Fishkin et al. (2024). In contrast, we are specifically interested in the outcomes separately. Nonetheless, it would be possible to apply similar analyses to our approach if desired. Fishkin et al. (2024) conduct an analysis over time. They use waves over a timeframe of one year, which is longer than the 3 months over which waves were conducted here. They test for any lasting effects over a period of 6 months. This is similar in essence to our measure of reversion, though we modelled the extent of reversion over a much shorter period. In this sense, perhaps our measure is better thought of as *short term reversion*. Reversion has also been studied within cognitive psychology through the study of models that specifically capture "preference reversal" (Johnson & Busemeyer, 2005; Tsetsos et al., 2010) though this work focussed on discrete choice rather than ratings data as we do here.

*Strengths and limitations*

Strengths of this study include, firstly, its size. The Food Conversation is the largest food-based deliberative event in the UK. Secondly, its representation. The data collection purposefully collected a balanced sample of a range of demographics to maximise representativeness. Importantly, all four countries of the UK. Third, a rich set of covariates were collected to allow analyses using these. Fourth, a strength was its design, specifically to have collected responses for five stakeholder groups, generating a larger dataset than would be available for having only collected data for a single stakeholder. Our approach, although based on an ordered outcome, can use any form of outcome variable (binary, multinomial, continuous, etc.) simply by adapting the nature of the question asked. In addition, our approach is very simple to implement. Fifth, whilst we used digital devices to collect data, other forms of data collection are perfectly possible, which could increase inclusion (e.g., collection with audio equipment or writing). Sixth, the marginal cost of implanting our approach is also very low. In its absence, the qualitative data collection and analyses would still have occurred in the same way. Here, a few minutes at the start and end of each session were used to collect data.

We are subject to a series of limitations. First, only a single question was asked (though asked across five stakeholders). This limits the qualitative analysis and interpretation to this question. This question was one of the main subjects of interest for the Food Conversation, but there were others. We therefore do not provide analyses of other aspects of this topic. It would have been possible to collect responses to more questions, though this increases the burden on respondents. Second, modelling relies on a relatively large dataset. Here, we had a large sample size and asked them to respond to five questions at ten time points, facilitating this data collection; this may not be possible elsewhere (e.g. smaller events would run only one or two workshops). Third, of the available data, only a subset could be used in modelling: of 345 individuals in the Food Conversation, 339 were available to us for modelling, and 213 of those could be used in analyses. These results are therefore vulnerable to forms of selection bias. For example, reports from the sessions suggested that older adults had some difficulties engaging with the digital devices for recording responses. Indeed, our sample mean age of 50 was below that which we calculated using ONS data of the average adult age in the UK of 56. Relatedly, there are high levels of missing demographic data within the sample: ranging from 24.9% (e.g., ethnicity) to 44.6% (e.g., IMD). This not only diminishes our ability to model variation across demographics, it could also induce non-response bias.

*Implications*

These methods are a low-cost, straightforward expansion of the toolkit available to researchers conducting deliberative research. They provide a quantitative complement to qualitative analysis, that may help to enrich study findings. Whilst we used ordered outcomes, there is no reason that binary, discrete, or continuous outcomes could be adopted. Our focus was on the food system, but these methods can be readily applied in any sector. We presented a modelling framework, with accompanying code online to facilitate use by other researchers.

*Conclusion and take-home point*

Qualitative analyses of deliberative events can be easily expanded with a complementary quantitative component.